\documentclass[pra,aps,preprint,amsmath,amssymb,showpacs]{revtex4}
\usepackage{graphicx}
\usepackage{epsfig}
\include{graphics}
\begin{document}

\title{Stable long-distance propagation and on-off switching  
of colliding soliton sequences with dissipative interaction}

\author{Debananda Chakraborty$^{1}$, Avner Peleg$^{2}$, 
Jae-Hun Jung$^{2}$}

\affiliation{$^{1}$ Department of Mathematics, Virginia Intermont College, 
Bristol, VA 24201, USA
\\
$^{2}$ Department of Mathematics, University at Buffalo, Buffalo, 
NY 14260, USA}

\date{\today}

\begin{abstract}
We study propagation and on-off switching of two 
colliding soliton sequences in the presence 
of second-order dispersion, Kerr nonlinearity, 
linear loss, cubic gain, and quintic loss.  
Employing a Lotka-Volterra (LV) model for dynamics 
of soliton amplitudes along with simulations with 
two perturbed coupled nonlinear Schr\"odinger (NLS) equations, 
we show that stable long-distance propagation 
can be achieved for a wide range of the gain-loss 
coefficients, including values that are outside of the perturbative regime. 
Furthermore, we demonstrate robust on-off and off-on switching 
of one of the sequences by an abrupt change in the ratio of cubic gain 
and quintic loss coefficients, and extend the results    
to pulse sequences with periodically alternating phases. 
Our study significantly strengthens the recently found relation 
between collision dynamics of sequences of NLS solitons 
and population dynamics in LV models, and indicates that 
the relation might be further extended to solitary waves 
of the cubic-quintic Ginzburg-Landau equation.        
\end{abstract}
\pacs{42.65.Tg, 05.45.Yv, 42.65.Sf}
\maketitle

\section{Introduction}
\label{Introduction}
The cubic nonlinear Schr\"odinger (NLS) equation, 
which describes wave propagation in the presence of 
second-order dispersion and cubic (Kerr) nonlinearity, 
is one of the most widely researched nonlinear wave models 
in physics. It was successfully employed to describe 
water wave dynamics \cite{Zakharov84}, 
nonlinear waves in plasma \cite{Horton96}, 
Bose-Einstein condensates (BECs) \cite{Dalfovo99}, 
and pulse propagation in optical waveguides \cite{Hasegawa95}.  
The fundamental NLS solitons are  
the most ubiquitous solutions of the cubic NLS equation 
due to their stability and to the fact that a generic 
wave pattern typically evolves into a sequence 
of fundamental solitons in the presence of 
anomalous dispersion and Kerr nonlinearity.
In the absence of additional physical 
processes (perturbations), the fundamental solitons propagate 
without any change in their amplitude, group velocity, and shape. 
However, the presence of perturbations usually breaks 
this ideal picture, by inducing changes in the solitons' 
amplitude, group velocity, and shape. 
Dissipative perturbations  due to linear and nonlinear 
gain or loss are of particular interest, since they are very 
common in soliton systems. In optical waveguides, for example,  
nonlinear loss or gain arise due to multiphoton absorption or 
emission, respectively \cite{Boyd2008}. 
Moreover, nonlinear loss and gain play an important role 
in many phenomena described by the complex 
Ginzburg-Landau (GL) equation \cite{Kramer2002}, 
such as convection and pattern formation in fluids 
and mode-locked lasers.

Despite its success, the single NLS equation is limited to 
describing a scalar physical field and is also unsuitable for 
handling generic broadband nonlinear wave systems. 
If the physical field is a vector, or if the spectra of the 
waves in a broadband wave system are concentrated about multiple 
widely separated wavelengths, the single NLS equation 
should be replaced by a system of coupled-NLS equations. 
Indeed, in recent years, coupled-NLS models have been 
employed in studies of a wide range of phenomena in fluid 
dynamics \cite{Fluids,Onorato2006}, 
nonlinear optics \cite{Hasegawa95,NLoptics}, 
multiphase BECs \cite{BECs}, and plasma physics \cite{plasma}. 
Coupled-NLS models are particularly useful in describing  
broadband wave systems, in which the waves are organized in 
sequences (trains) moving with very different group velocities, 
such as in crossing seas \cite{Onorato2006}, 
or in broadband optical waveguide transmission \cite{Hasegawa95}. 
Due to the large group velocity differences 
in these systems, collisions between pulses from different 
sequences are very frequent, and thus play a major role in 
the dynamics.

In the current paper, we study the dynamics 
of two colliding sequences of fundamental NLS solitons 
in the presence of dissipative perturbations due to  
linear loss, cubic gain, and quintic loss 
(i.e., a GL gain-loss profile). In the absence of perturbations, 
the solitons' amplitudes and group velocities do not 
change in the collisions. However, the presence of 
dissipative interaction due to cubic gain and quintic loss 
induces additional amplitude shifts during the collisions, 
whose magnitude depends on the initial amplitudes of the two 
colliding solitons. Consequently, amplitude dynamics of 
solitons in the two sequences become nonlinearly coupled,  
and important questions arise regarding the nature of 
this dynamics. In three recent studies we showed that 
amplitude dynamics of $N$ sequences of colliding solitons  
in the presence of dissipative perturbations might be 
described by Lotka-Volterra (LV) models for $N$ species, 
where the exact form of the LV model depends 
on the nature of the perturbations \cite{NP2010a,PNC2010,PC2012b}.       
This potential relation between coupled-NLS equations and LV 
systems is of great interest, due to the central role 
of these seemingly unrelated models in the natural sciences. 
While the importance of coupled-NLS equations was explained 
in the previous paragraph, LV models are widely used in 
environmental sciences to describe population dynamics 
\cite{Lotka25,Volterra28,LV_ecology}, in chemistry, to study 
dynamics of chemical reactions \cite{LV_chem}, in economics, 
to explain interaction between different technologies
\cite{LV_economics}, and in neural networks, to compute neuron 
firing rates \cite{LV_nueral}. Due to the importance 
of coupled-NLS and LV models, it is essential to further enhance 
and extend the understanding of the relation between them. 
Moreover, a better understanding of this relation can be used 
for stabilizing the propagation, for controlling and tuning 
of soliton amplitudes and group velocities, and even for 
broadband switching.

In order to give further motivation for the current study 
and to clarify the importance of its results, we provide a 
brief critical description of our three earlier studies of 
the problem. In Ref. \cite{NP2010a}, we studied the propagation 
of colliding soliton sequences in the presence of delayed Raman 
response. We showed that amplitude dynamics of $N$ soliton 
sequences is described by an $N$-dimensional 
predator-prey model, but the results were not checked 
by numerical simulations with the corresponding coupled-NLS 
model. Furthermore, the equilibrium states of the predator-prey 
model were found to be centers, i.e., the equilibria were not 
asymptotically stable. In order to resolve the problem of 
asymptotic stability in {\it the LV model}, 
we turned to investigate collision dynamics 
of soliton sequences in the presence of weak linear gain 
and cubic loss \cite{PNC2010}. We showed that in this 
case dynamics of soliton amplitudes might be approximately 
described by a LV model for competing species with quadratic 
interaction terms. Stability analysis of the equilibrium 
states of the model yielded conditions on the 
physical parameters for stable propagation of two sequences 
with equal soliton amplitudes. Numerical simulations with the full 
coupled-NLS model confirmed stability at short-to-intermediate 
distances, but uncovered an instability at longer 
distances \cite{PNC2010}. The latter finding was attributed 
to weak (second-order) radiative instability due to the impact 
of linear gain on the two sequences. In order to overcome this 
destabilizing effect, we turned to study a different setup, 
where the soliton sequences propagate in the presence of weak 
linear loss, cubic gain, and quintic loss \cite{PC2012b}.
We showed that in this setup, the LV model for amplitude dynamics   
contains quadratic and quartic interaction terms. The presence of linear 
loss was expected to suppress the weak radiative instability 
and by this to enable stable long-distance propagation. 
However, numerical solution of the corresponding 
perturbed coupled-NLS model still showed instability 
at intermediate-to-long propagation distances \cite{PC2012b}. 
Thus, the important problem of further stabilization 
of the propagation remained unresolved. 
Moreover, due to the instability at intermediate 
distances, all the previous studies were unable to explore  
the possibility of achieving stable on-off and off-on 
switching, i.e., the turning on and off of transmission 
of one sequence by a fast change of one or more of the 
physical parameters. In addition, the earlier studies were 
limited in scope to small values of the dissipative coefficients 
and to uniform initial distribution of soliton amplitudes 
and phases in each sequence.

In the current paper we address these important deficiencies 
of the earlier studies. Considering propagation of two colliding 
soliton sequences in the presence of a GL gain-loss profile, 
we observe stable long-distance transmission over a wide range 
of the gain-loss coefficients, including values that are outside 
of the perturbative regime. The stable transmission is enabled 
by the presence of linear loss, a sufficiently large initial 
inter-soliton separation, and the absence of significant 
initial position shift between the two soliton sequences.  
Moreover, we show that robust on-off and off-on switching 
of one soliton sequence can be realized by an abrupt change 
in the ratio of the cubic gain and quintic loss coefficients. 
In addition, we extend the results to soliton sequences with 
periodically alternating phases. Our findings significantly 
enhance the surprising relation between propagation of 
colliding NLS soliton sequences and population dynamics 
in LV models. The results also indicate that the relation 
might be further extended to sequences of solitons of 
the cubic-quintic GL equation.

The rest of the paper is organized as follows. 
In Sec. \ref{models}, we present the coupled-NLS model 
for the propagation along with the reduced LV model 
for amplitude dynamics. We also discuss the predictions 
of the latter model for stability of propagation. 
In Sec. \ref{simu}, we present the results of numerical 
simulations with the coupled-NLS model for long distance 
transmission, as well as for on-off and off-on switching.  
In addition, we compare the simulation results 
with the predictions of the LV model. 
Our conclusions are presented in 
Sec. \ref{conclusions}.

\section{Coupled-NLS and Lotka-Volterra models}
\label{models} 

We consider propagation of sequences of soliton pulses in 
the presence of second-order dispersion, Kerr nonlinearity, 
and a GL gain-loss profile, consisting of 
linear loss or gain, cubic gain, and quintic loss. 
We denote the physical field of the $j$th sequence by $\psi_{j}$. 
In the context of propagation of light through optical waveguides, 
for example,  $\psi_{j}$ is proportional to 
the envelope of the electric field of the $j$th sequence. 
In the absence of gain and loss, the propagation 
of the $j$th soliton-sequence is described by the 
cubic NLS equation
\begin{eqnarray} &&
i\partial_z\psi_{j}+\partial_{t}^2\psi_{j}+2|\psi_{j}|^2\psi_{j}=0,
\label{crosstalk1}
\end{eqnarray}        
where we adopt the waveguide optics notation, in which 
$z$ is propagation distance, and $t$ is time. 
The fundamental soliton solution of the NLS equation  
with group velocity $2\beta_{j}$ 
is  $\psi_{sj}(t,z)=\eta_{j}\exp(i\chi_{j})\mbox{sech}(x_{j})$,
where $x_{j}=\eta_{j}\left(t-y_{j}-2\beta_{j} z\right)$, 
$\chi_{j}=\alpha_{j}+\beta_{j}(t-y_{j})+
\left(\eta_{j}^2-\beta_{j}^{2}\right)z$, 
and $\eta_{j}$, $y_{j}$, and $\alpha_{j}$ 
are the soliton amplitude, position, and phase, respectively.

We focus attention on the dynamics of two sequences of fundamental 
NLS solitons propagating with group velocities $2\beta_{j}$, 
where $j=1,2$. Assuming a large group velocity difference 
$|\beta_{1}-\beta_{2}|\gg 1$, the solitons undergo a large number 
of fast inter-sequence collisions. We now take into  account 
the effects of linear gain-loss, cubic gain, and quintic loss   
as well as inter-sequence interaction due to Kerr nonlinearity.     
Thus, the propagation is described by the following system 
of perturbed coupled-NLS equations: 
\begin{eqnarray} &&
i\partial_z\psi_{j}+\partial_{t}^2\psi_{j}+2|\psi_{j}|^2\psi_{j}
+4|\psi_{k}|^2\psi_{j}=
\nonumber \\&&
ig_{j}\psi_{j}/2
+i\epsilon_{3}|\psi_{j}|^2\psi_{j}+2i\epsilon_{3}|\psi_{k}|^2\psi_{j}
\nonumber \\&&
-i\epsilon_{5}|\psi_{j}|^4\psi_{j}-3i\epsilon_{5}|\psi_{k}|^4\psi_{j}
-6i\epsilon_{5}|\psi_{k}|^2|\psi_{j}|^2\psi_{j},
\label{crosstalk2}
\end{eqnarray}          
where $j=1,2$, $k=1,2$, $g_{j}$ is the linear 
gain-loss coefficient for the $j$th sequence, 
and $\epsilon_{3}$ and $\epsilon_{5}$ are the cubic gain 
and quintic loss coefficients, respectively.     
The term $4|\psi_{k}|^2\psi_{j}$ in Eq. (\ref{crosstalk2}) 
describes inter-sequence interaction due to Kerr nonlinearity, 
while $ig_{j}\psi_{j}/2$, $i\epsilon_{3}|\psi_{j}|^2\psi_{j}$, and 
$-i\epsilon_{5}|\psi_{j}|^4\psi_{j}$ correspond to 
intra-sequence effects due to linear gain-loss, cubic gain and quintic loss, 
respectively. In addition, the term $2i\epsilon_{3}|\psi_{k}|^2\psi_{j}$ 
represents inter-sequence effects due to cubic gain, while 
$-3i\epsilon_{5}|\psi_{k}|^4\psi_{j}$ and    
$-6i\epsilon_{5}|\psi_{k}|^2|\psi_{j}|^2\psi_{j}$ 
describe dissipative inter-sequence interaction due to quintic loss.

In Ref. \cite{PC2012b}, we showed that under certain assumptions, 
amplitude dynamics of solitons in the two sequences can be approximately 
described by a LV model for two species with quadratic 
and quartic interaction terms. The derivation of the LV model was based 
on the following assumptions. 
(1) The temporal separation $T$ between 
adjacent solitons in each sequence is a large constant: $T \gg 1$. 
In addition, the amplitudes are equal for all solitons from the 
same sequence, but are not necessarily equal for solitons from 
different sequences. (2) The pulses circulate in a closed 
loop, e.g., in an optical waveguide ring.  
(3) As $T\gg 1$, the pulses in each sequence are 
temporally well-separated. As a result, intra-sequence interaction is 
exponentially small and is neglected. (4) High-order effects due to 
collision-induced frequency shift and emission of radiation 
are also neglected.

Since the soliton sequences are periodic, 
the amplitudes of all pulses in a given sequence undergo the same 
dynamic evolution. Taking into account collision-induced 
and single-pulse amplitude changes, we obtain the following equation 
for the rate of change of the amplitude of the solitons in the $j$th 
sequence $\eta_{j}$ \cite{PC2012b}: 
\begin{eqnarray} &&
\frac{d \eta_{j}}{dz}=
\eta_{j}\left[g_{j}+\frac{4}{3}\epsilon_{3}\eta_{j}^{2}
-\frac{16}{15}\epsilon_{5}\eta_{j}^{4}
+\frac{8}{T}\epsilon_{3}\eta_{k}
-\frac{8}{T}\epsilon_{5}\eta_{k}\left(2\eta_{j}^{2}+\eta_{k}^{2}\right)
\right],  
\label{crosstalk3}
\end{eqnarray}
where $j=1,2$. Note that Eq. (\ref{crosstalk3}) can be described as 
a LV model for two species with quadratic cooperation terms   
and quartic competition terms.

The results described in the previous paragraph indicate that 
there is a relation between collision-induced dynamics of soliton 
sequences in nonlinear waveguides with a GL gain-loss profile 
and dynamics of population size of species with nonlinear 
competition and cooperation. Alternatively, one may speak 
about a connection between the 
perturbed coupled-NLS model (\ref{crosstalk2}) 
and the LV model (\ref{crosstalk3}). This relation  
along with knowledge about properties of LV models
can be used to develop ways for controlling the dynamics 
of the colliding solitons. A straightforward way for 
achieving this goal is by tuning of the 
linear gain-loss coefficients $g_{j}$. In optical waveguide 
systems, for example, this can be realized  by adjusting 
the linear amplifier gain.

As a particular example, we consider the important case where the 
values of the $g_{j}$ coefficients are chosen such that the LV model 
(\ref{crosstalk3}) has a steady state $(\eta,\eta)$ with equal 
amplitudes for both sequences. This case is of special interest, 
since in optical waveguide systems it is usually easier to control 
the transmission when the amplitudes of all pulses are equal.
Requiring that $(\eta,\eta)$ is a steady state, we obtain
$g_{j}=4\epsilon_{5}\eta\left(-\kappa\eta/3+4\eta^{3}/15
-2\kappa/T+6\eta^{2}/T\right)$, where 
$\kappa=\epsilon_{3}/\epsilon_{5}$ and $\epsilon_{5}\ne 0$.    
Substituting this relation into Eq. (\ref{crosstalk3}),    
we arrive at the following LV model for amplitude dynamics:  
\begin{eqnarray} &&
\frac{d \eta_{j}}{dz}=
\epsilon_{5}\eta_{j}
\left\{\frac{4\kappa}{3}(\eta_{j}^{2}-\eta^{2}) 
-\frac{16}{15}(\eta_{j}^{4}-\eta^{4})
+\frac{8\kappa}{T}(\eta_{k}-\eta)
-\frac{8}{T}\left[\eta_{k}\left(2\eta_{j}^{2}+\eta_{k}^{2}\right)
-3\eta^{3}\right]\right\}.
\label{crosstalk4}
\end{eqnarray}     
We note that  $(\eta,\eta)$ and $(0,0)$ are equilibrium points 
of Eq. (\ref{crosstalk4}) for any positive values of $\eta$, 
$\kappa$, and $T$.

Let us discuss the predictions of the LV model (\ref{crosstalk4}) 
regarding stability of propagation of the two soliton sequences. 
For concreteness, we choose $\eta=1$, and require that the 
equilibrium points at $(1,1)$ and $(0,0)$ are stable nodes (sinks). 
The stability requirement on $(0,0)$, which is important 
for suppression of weak instability due to radiation emission,     
leads to $g_{j}<0$ for $j=1,2$, i.e., the solitons are subject to 
net linear loss. The above stability requirements yield the 
following conditions on $T$ and $\kappa$ \cite{PC2012b}:
$(4T+90)/(5T+30)<\kappa<(8T+135)/(5T+15)$ for $3 < T < 60/17$, 
and $(4T+90)/(5T+30)<\kappa<(8T-15)/(5T-15)$ for $T \ge 60/17$.

\section{Numerical simulations with the coupled-NLS model}
\label{simu}
As explained above, the LV model (\ref{crosstalk4}) provides 
an approximate description for collision-induced 
dynamics of circulating soliton sequences. 
The propagation is fully described by the coupled-NLS system 
(\ref{crosstalk2}) with periodic boundary conditions. 
A key question about the collision-induced dynamics 
concerns the possibility to suppress the instability 
observed in numerical simulations in Ref. \cite{PC2012b}. 
Here we address this important question. 
More specifically, we show that the instability can be effectively 
mitigated by using a larger temporal separation value T, compared 
with the one used in \cite{PC2012b}, and by eliminating the 
initial inter-sequence position-shift, which was introduced in 
\cite{PC2012b} to avoid incomplete collisions. 
Thus, the numerical simulations setup in the current paper 
consists of the coupled-NLS model (\ref{crosstalk2}) with 
periodic boundary conditions and an initial condition in the form of 
two periodic sequences of $2J+1$ {\it overlapping} solitons
with amplitudes $\eta_{j}(0)$ and zero phase:  
\begin{eqnarray} &&
\psi_{j}(t,0)\!=\!\sum_{k=-J}^{J}
\frac{\eta_{j}(0)\exp[i\beta_{j}(t-kT)]}
{\cosh[\eta_{j}(0)(t-kT)]}, 
\label{crosstalk5}
\end{eqnarray}
where $j=1,2$. The temporal separation $T$ is taken as $T=20$, 
compared with $T=10$ that was used in Ref. \cite{PC2012b}. 
Four different values of $\epsilon_{5}$ 
are used: $\epsilon_{5}=0.01$, $\epsilon_{5}=0.06$, 
$\epsilon_{5}=0.1$, and $\epsilon_{5}=0.5$, 
compared with $\epsilon_{5}=0.01$ that was 
considered in Ref. \cite{PC2012b}.       
Note that a value of $\epsilon_{5}=0.5$ is outside 
of the parameter range, where the perturbative approach 
leading to the LV model (\ref{crosstalk3}) is expected to be valid.  
The other physical parameters in the simulations 
are taken as  $\kappa=1.5$, $\eta=1$, $\beta_{1}=0$, 
$\beta_{2}=40$, and $J=2$.

\begin{figure}[ptb]
\begin{tabular}{cc}
\epsfxsize=6.0cm  \epsffile{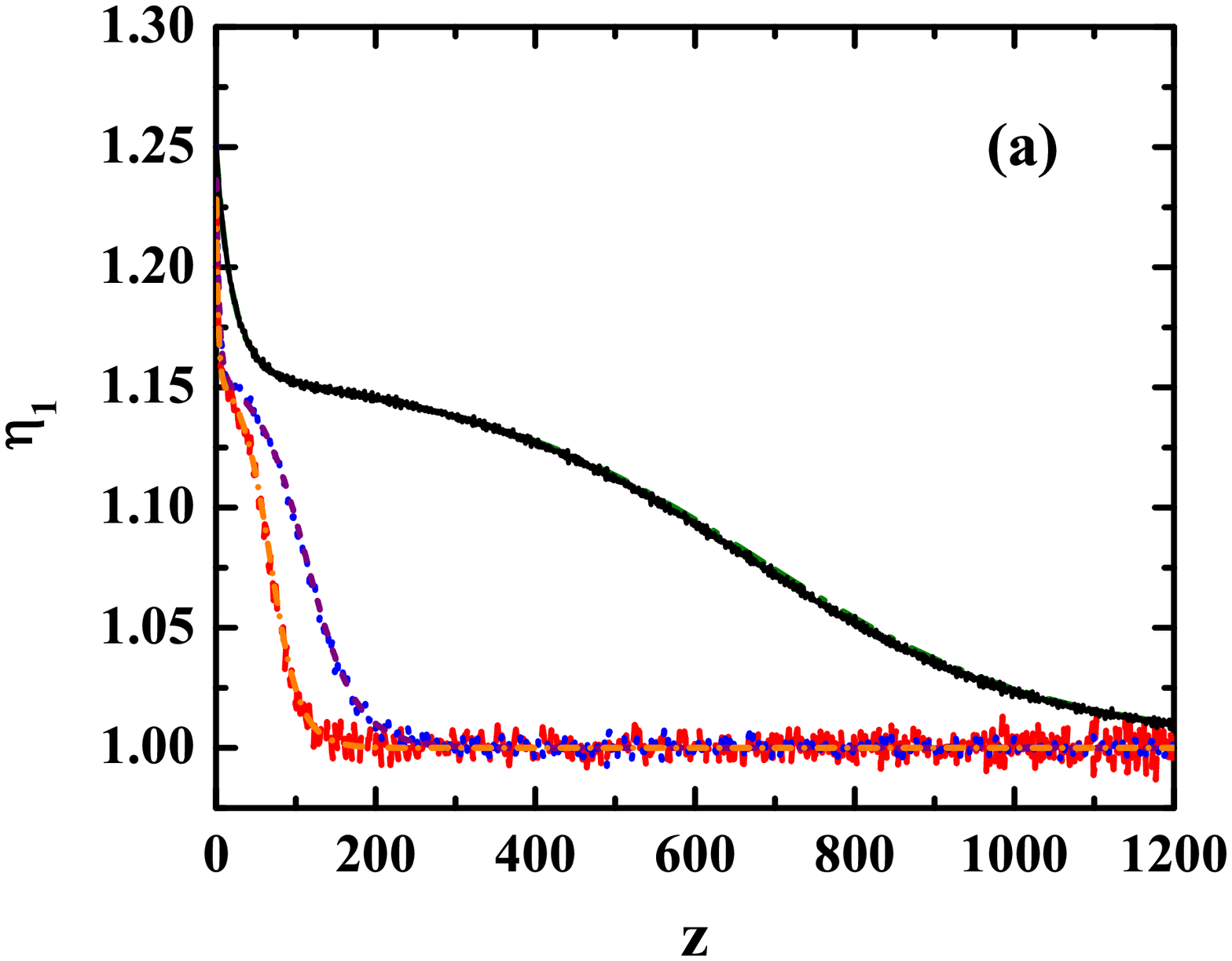} \\
\epsfxsize=6.0cm  \epsffile{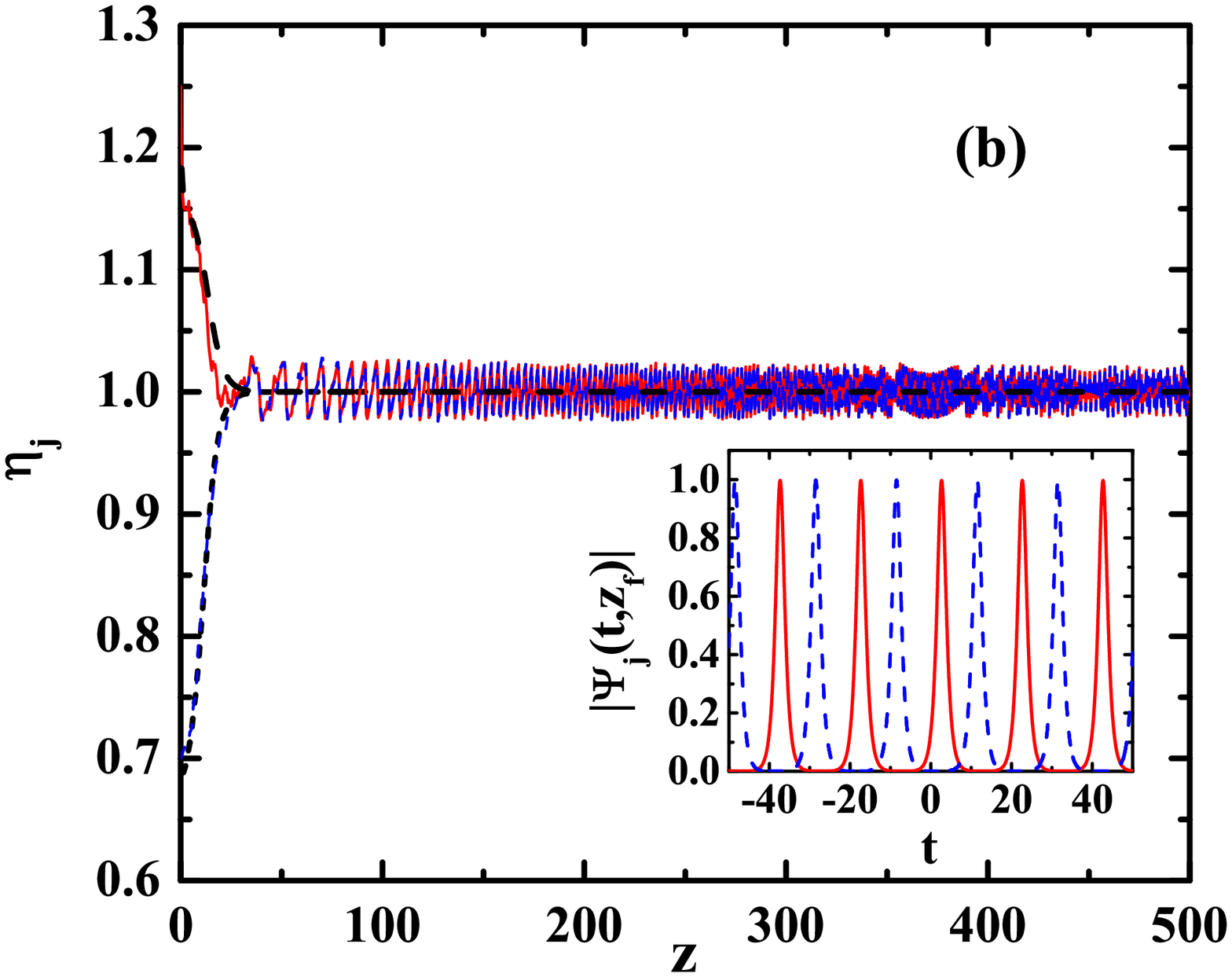}  
\end{tabular}
\caption{(Color online) 
(a) The $z$-dependence of $\eta_{1}$ for various $\epsilon_{5}$ values 
and $T=20$, $\kappa=1.5$, $\eta_{1}(0)=1.25$, and $\eta_{2}(0)=0.7$. 
The solid black, dotted blue, and dashed red lines correspond to    
$\eta_{1}(z)$ values obtained by numerical solution of the 
coupled-NLS model (\ref{crosstalk2}) with $\epsilon_{5}=0.01$, 
$\epsilon_{5}=0.06$, and $\epsilon_{5}=0.1$, respectively. 
The dashed-dotted green, short dashed purple, and short dashed-dotted 
orange lines represent $\eta_{1}(z)$ values predicted by the LV model 
(\ref{crosstalk4}) with  $\epsilon_{5}=0.01$, $\epsilon_{5}=0.06$, 
and $\epsilon_{5}=0.1$.        
(b) The z-dependence of $\eta_{1}$ and $\eta_{2}$ for $\epsilon_{5}=0.5$ 
and the same values of $T$, $\kappa$, $\eta_{1}(0)$, and $\eta_{2}(0)$ 
as in (a). The solid red and dashed blue lines represent   
$\eta_{1}(z)$ and $\eta_{2}(z)$ as obtained by numerical solution of 
Eq. (\ref{crosstalk2}), while the dashed and short dashed black lines 
correspond to $\eta_{1}(z)$ and $\eta_{2}(z)$ values as obtained by 
the LV model (\ref{crosstalk4}). The inset shows the final pulse 
patterns $|\psi_{1}(t,z_{f})|$ (solid red) and $|\psi_{2}(t,z_{f})|$ 
(dashed blue) obtained in the coupled-NLS simulation.} 
\label{fig1}
\end{figure}

The $z$-dependence of $\eta_{1}$ obtained by numerical 
solution of Eq. (\ref{crosstalk2}) with initial amplitudes 
$\eta_{1}(0)=1.25$ and $\eta_{2}(0)=0.7$ is shown in 
Fig. \ref{fig1}(a) for  $\epsilon_{5}=0.01$, $\epsilon_{5}=0.06$, 
and $\epsilon_{5}=0.1$, and in Fig. \ref{fig1}(b) for $\epsilon_{5}=0.5$.  
The predictions of the LV model (\ref{crosstalk4}) are also presented. 
The agreement between the coupled-NLS simulations and the predictions 
of the LV model are excellent. Moreover, in all 
four cases $\eta_{1}(z)$ tends to 
the equilibrium value of $\eta=1$ and the approach to 
$\eta=1$ is faster as $\epsilon_{5}$ increases. 
Similar results are obtained for $\eta_{2}(z)$ and 
for other initial amplitude values.  
Note that the distances over which stable propagation is 
observed are larger by a factor of 6.67 compared with the 
the distances in Ref. \cite{PNC2010}, and by a factor of 2 
or more compared with the distances in Ref. \cite{PC2012b}. 
The fact that the predictions of the LV model hold even 
for $\epsilon_{5}$ values as large as 0.5, i.e., outside 
of the perturbative regime, is quite surprising. 
Furthermore, as shown in the inset of Fig. \ref{fig1}(b), 
the shape of the two soliton trains is retained during 
the propagation despite of the magnitude of $\epsilon_{5}$ 
and the large number of collisions.

The observation of long-distance propagation of the two 
soliton sequences opens the way for studying broadband 
on-off and off-on switching, i.e., the turning on and 
off of one of the propagating pulse sequences. The two 
switching scenarios are based on bifurcations of the 
steady state $(1,1)$ of the LV model. More specifically, 
in on-off switching, the value of the parameter $\kappa$ 
is abruptly increased at the switching distance $z_{s}$ 
from $\kappa_{i}<\kappa_{c}$ to $\kappa_{f}>\kappa_{c}$, 
where $\kappa_{c}=(8T-15)/(5T-15)$,   
such that the steady state $(1,1)$ becomes unstable, 
while another steady state at $(\eta_{s},0)$ is stable. 
As a result, soliton amplitudes in sequences 2 and 1 
tend to 1 for $z<z_{s}$, but tend to 0 and $\eta_{s}$, 
respectively, for $z>z_{s}$. This means that transmission 
of sequence 2 is effectively turned off at $z_{s}$.       
Off-on switching is realized in a similar manner, 
by decreasing the value of $\kappa$ at the switching distance 
$z_{s}$ from $\kappa_{i}>\kappa_{c}$ to $\kappa_{f}<\kappa_{c}$  
such that the steady state $(1,1)$ becomes stable. 
Consequently, for $z<z_{s}$, the amplitudes of solitons in sequences 
1 and 2 tend to $\eta_{s}$ and 0, respectively, 
but for $z>z_{s}$, both amplitudes tend to 1.         
Thus, in this case transmission of sequence 2 is turned 
on at $z_{s}$.

\begin{figure}[ptb]
\begin{tabular}{cc}
\epsfxsize=6.0cm  \epsffile{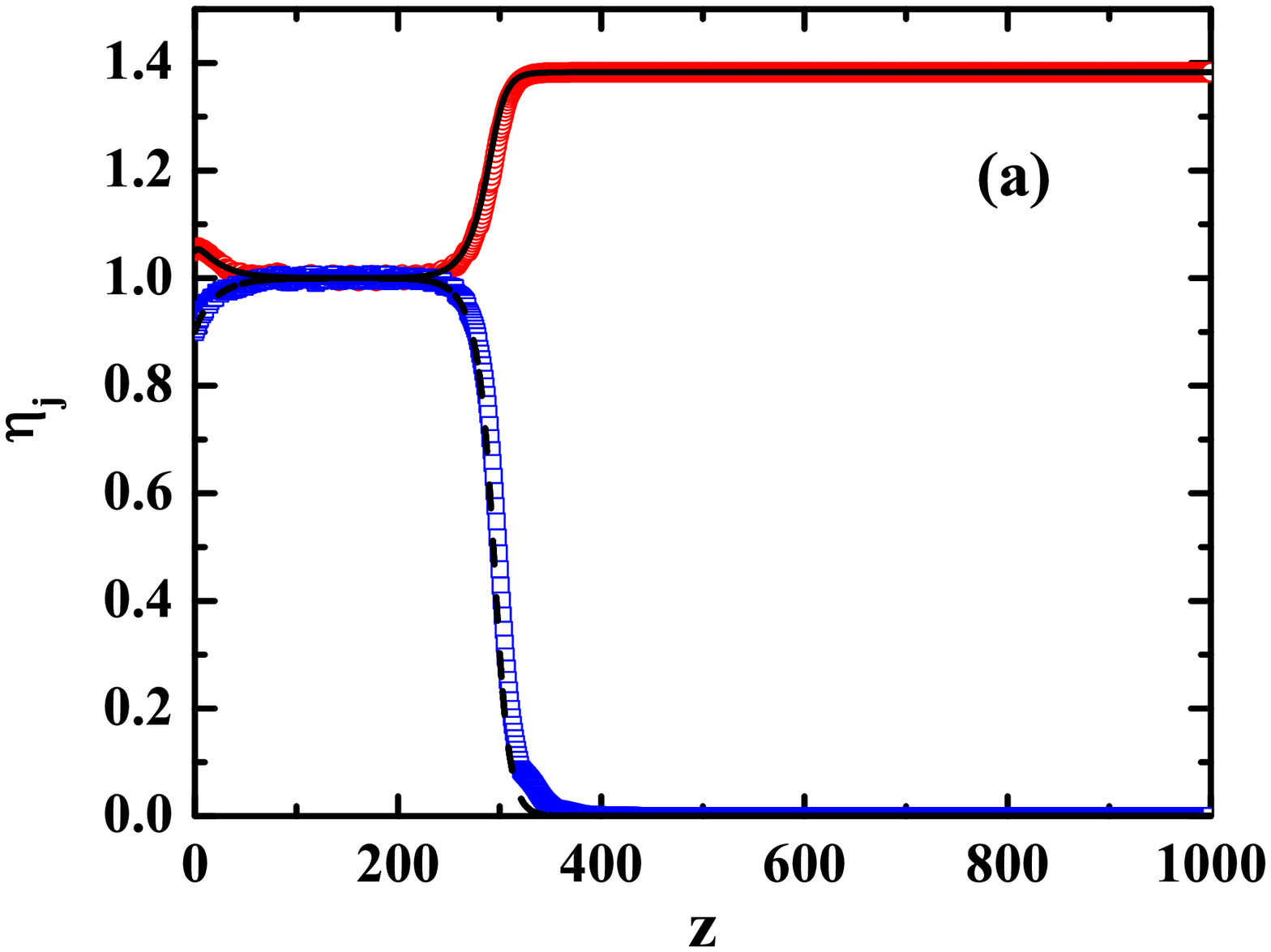} \\
\epsfxsize=6.0cm  \epsffile{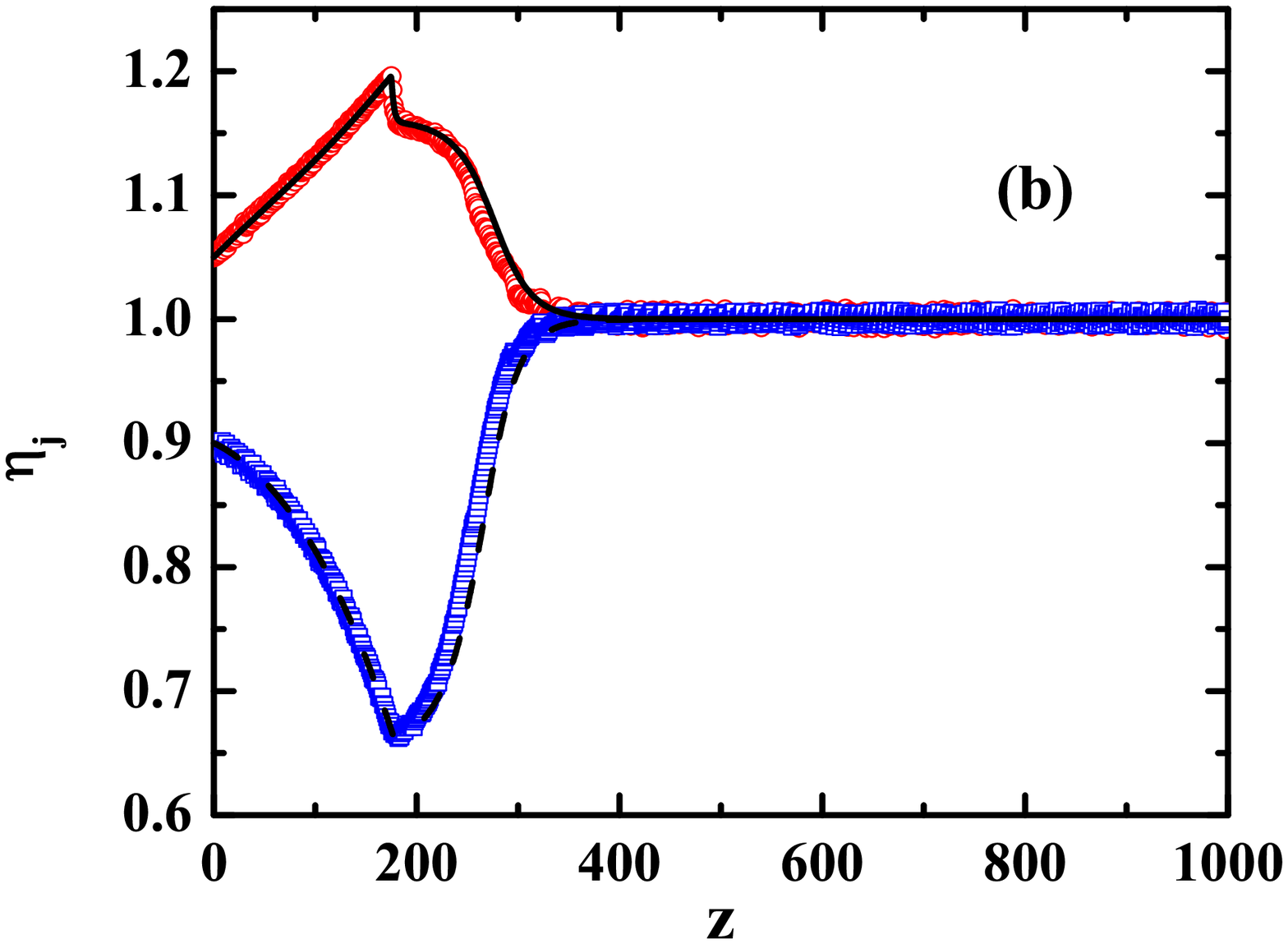} 
\end{tabular}
\caption{(Color online) 
The $z$-dependence of soliton amplitudes in on-off switching (a), 
and in off-on switching (b) with $z_{s}=175$ and parameter values 
as described in the text. The red circles and blue squares 
represent $\eta_{1}(z)$ and $\eta_{2}(z)$ values obtained 
with the coupled-NLS model (\ref{crosstalk2}), 
while the solid and dashed black lines correspond to 
$\eta_{1}(z)$ and $\eta_{2}(z)$ values obtained with the 
LV model (\ref{crosstalk4}).} 
\label{fig2}
\end{figure}

In order to check if these switching scenarios can be realized 
with sequences of colliding solitons, we numerically solve  
Eq. (\ref{crosstalk2}) with initial pulse patterns of the 
form (\ref{crosstalk5}), where $T=20$, $\eta_{1}(0)=1.05$, 
and $\eta_{2}(0)=0.9$. Since $T=20$, the critical 
$\kappa$-value is $\kappa_{c}=29/17$.    
As an example, the switching distance is taken as $z_{s}=175$. 
In on-off switching, $\epsilon_{5}=0.1$, and $\kappa$ is 
increased from $1.5$ to $2.0$ at $z_{s}$. 
In off-on switching, we use $\epsilon_{5}=0.01$ 
and $\kappa=2.0$ for $z \le 175$, and $\epsilon_{5}=0.1$ 
and $\kappa=1.5$ for $z > 175$. The results of the 
simulations for $\eta_{j}(z)$ are shown in 
Fig. \ref{fig2} (a) and (b) for on-off and off-on switching, 
respectively. Also shown is the prediction of the LV 
model (\ref{crosstalk4}). The overall agreement 
between the coupled-NLS simulations and 
the predictions of the LV model is very good for both 
switching scenarios. The only exception is for small values 
of $\eta_{2}(z)$ in on-off switching, where the LV model 
underestimates the value obtained with the coupled-NLS model. 
This is due to the fact that in this regime, the linear loss 
term is comparable to or larger than the Kerr nonlinearity term, 
leading to the breakdown of the perturbative description. 
Despite of this fact, the amplitudes of the solitons 
in sequence 2 do tend to 0, and thus, 
the on-off switching is fully realized.       
Based on the good agreement between coupled-NLS simulations 
and predictions of the LV model we conclude that on-off and 
off-on switching can be realized for the two sequences 
of colliding solitons. It should be noted that stable 
off-on switching can be achieved only when 
$\eta_{2}(z_{s})$ is larger than some threshold 
value $\eta_{th}$. For the parameter values used in our 
simulations, $\eta_{th}=0.65$. Thus, in order to implement  
off-on switching with the current setup of two soliton sequences, 
the value of the decision level distinguishing between 0 and 1 
states should be set larger than $\eta_{th}$.

\begin{figure}[ptb]
\begin{tabular}{cc}
\epsfxsize=6.0cm  \epsffile{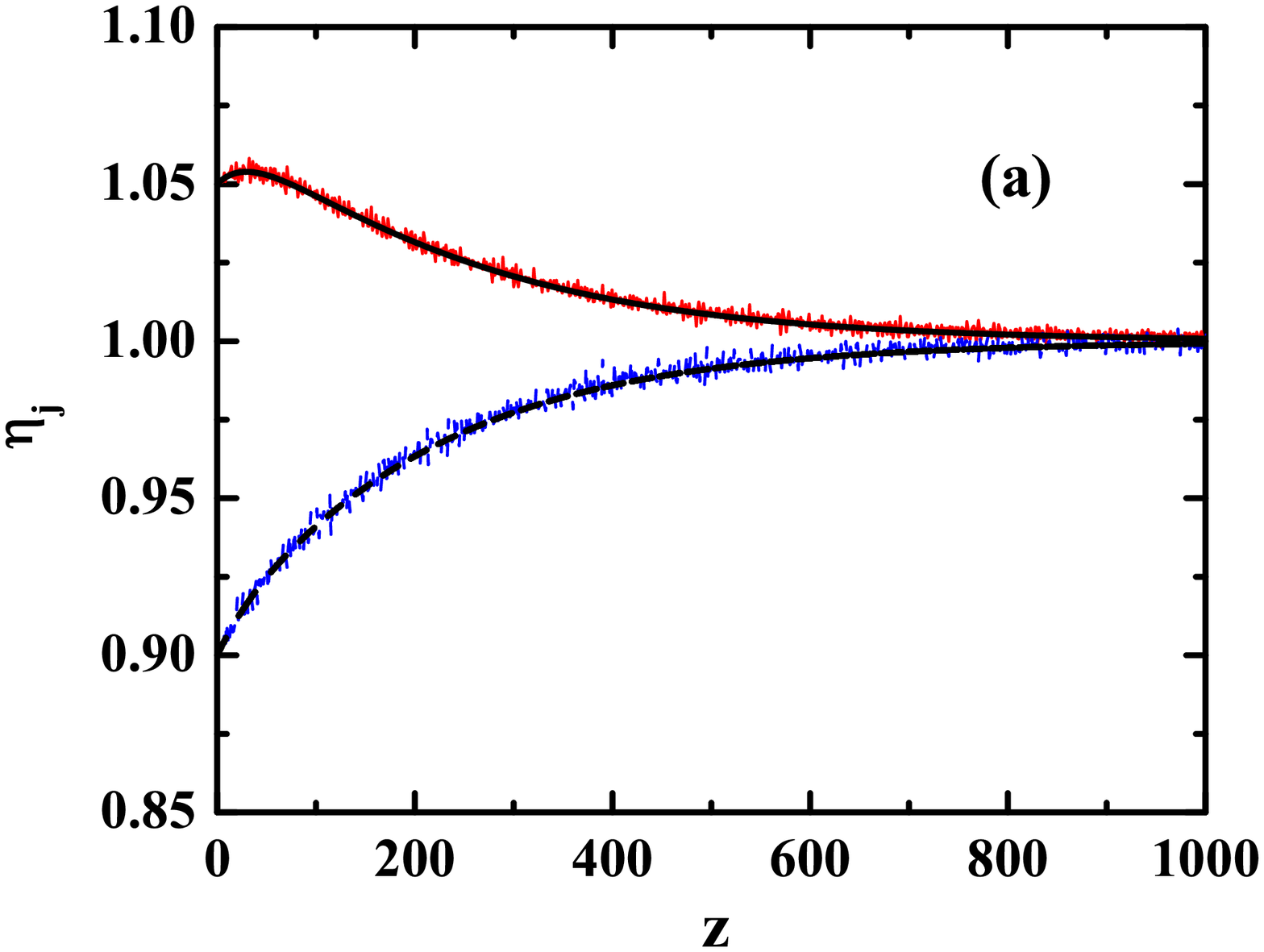}  \\
\epsfxsize=6.0cm  \epsffile{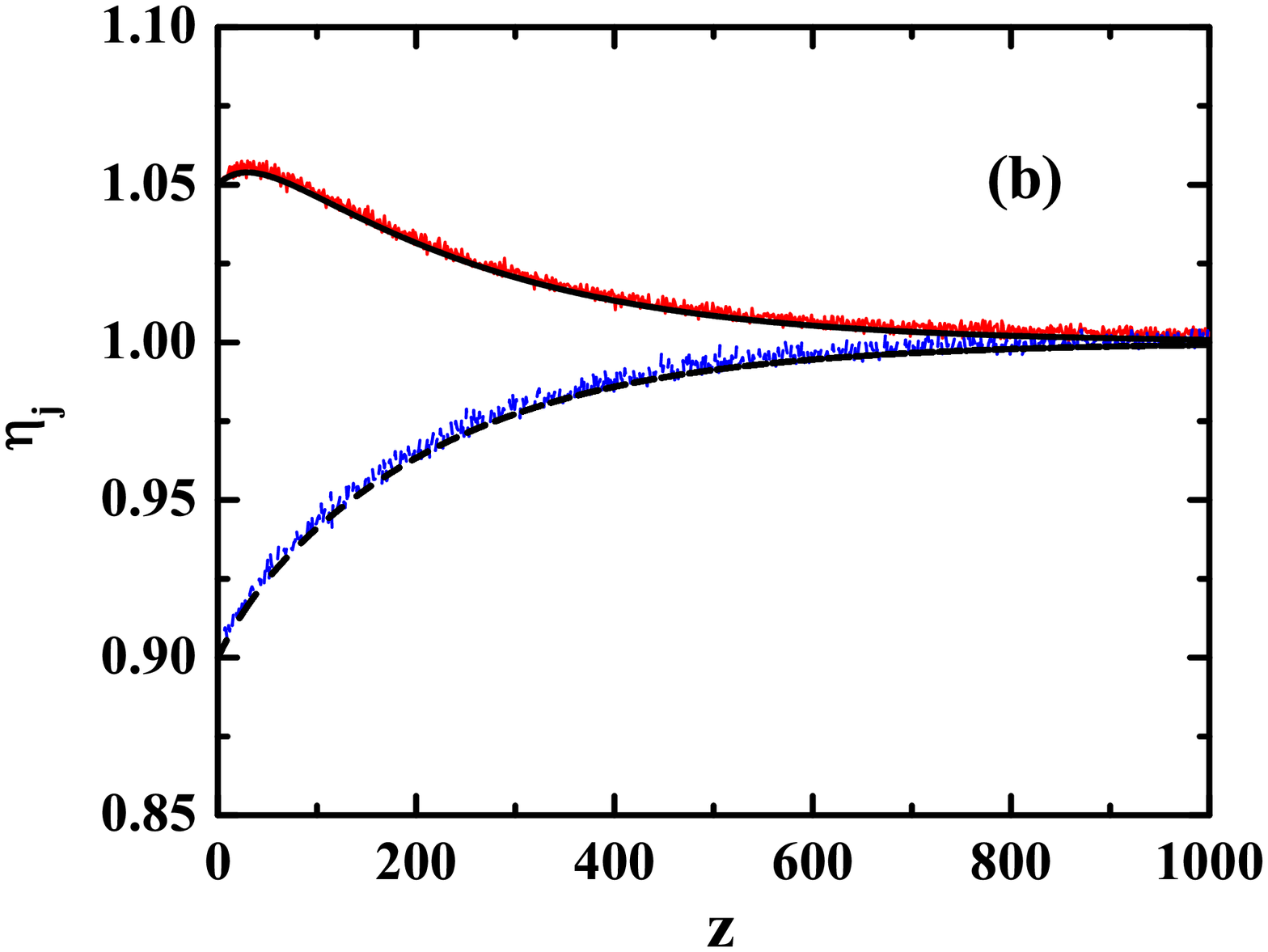}
\end{tabular}
\caption{(Color online) 
The $z$-dependence of pulse amplitudes for soliton sequences 
with alternating initial phases and parameter values $T=20$, 
$\kappa=1.5$, $\epsilon_{5}=0.01$, $\eta_{1}(0)=1.05$, 
and $\eta_{2}(0)=0.9$. (a) $\eta_{j}$ vs $z$ for    
$\alpha_{jk}(0)=0$, $\alpha_{jk+1}(0)=\pi$. 
(b) $\eta_{j}$ vs $z$ for $\alpha_{jk}(0)=0$, $\alpha_{jk+1}(0)=\pi/2$.
The solid red and dashed blue lines represent 
$\eta_{1}(z)$ and $\eta_{2}(z)$ values obtained 
with the coupled-NLS model (\ref{crosstalk2}), 
while the solid and dashed black lines correspond to 
$\eta_{1}(z)$ and $\eta_{2}(z)$ values obtained with the 
LV model (\ref{crosstalk4}).} 
\label{fig3}
\end{figure}

All the propagation setups discussed so far are limited, 
in the sense that all amplitudes and phases within each soliton 
sequence are equal. It is therefore important to extend the results 
to more general setups. Possible extension can be achieved by launching 
soliton sequences with periodically alternating phases. 
Indeed, the LV model (\ref{crosstalk4}) is independent 
of the soliton phases, and as a result, 
amplitude dynamics for sequences with periodically 
alternating phases is expected to be the same as the 
dynamics for soliton sequences with a uniform phase pattern. 
On the other hand, it is known that intra-sequence soliton 
interaction strongly depends on the relative phase 
difference between the solitons. The interplay between this 
interaction and other effects, such as loss or radiation 
emission might induce instabilities, which would lead to 
the breakdown of the LV model description. 
For this reason, it is important to check whether the predictions 
of the LV model  (\ref{crosstalk4}) do apply for soliton 
sequences with periodically alternating phases. 
For this purpose, we numerically solve 
Eq. (\ref{crosstalk2}) with pulse sequences consisting of two 
solitons each with initial amplitudes $\eta_{1}(0)=1.05$ (sequence 1) 
and $\eta_{2}(0)=0.9$ (sequence 2). 
Two sets of initial phases are used: 
$\alpha_{j1}(0)=0$ and $\alpha_{j2}(0)=\pi$ [set(a)]; 
and $\alpha_{j1}(0)=0$ and $\alpha_{j2}(0)=\pi/2$ [set(b)]. 
The other physical parameter values are taken as 
$T=20$, $\kappa=1.5$, and $\epsilon_{5}=0.01$. 
The results of the numerical simulations for 
$\eta_{j}(z)$ are shown in Fig. \ref{fig3} 
along with the predictions of the LV model. 
As can be seen, for both sets of initial phases 
the agreement between full-scale simulations 
and the predictions of the LV model is excellent 
over the entire propagation distance.  
Furthermore, the soliton amplitudes tend to the 
equilibrium value $\eta=1$, i.e., the propagation 
is indeed linearly stable. Thus, our numerical 
simulations demonstrate that the predictions of 
the LV model (\ref{crosstalk4}) can be applied to 
generic setups with periodic phase patterns.

\section{Conclusions}
\label{conclusions}
In summary, we investigated long-distance propagation 
as well as on-off and off-on switching 
of two colliding soliton sequences in the presence 
of second-order dispersion, Kerr nonlinearity, and a GL gain-loss 
profile. Using coupled-NLS simulations and a reduced LV model,  
we showed that stable long-distance propagation 
is possible for a wide range of the gain-loss coefficients, 
including values that are outside of the perturbative regime. 
The stable propagation is enabled by the presence of linear loss, 
the choice of a sufficiently large initial inter-pulse 
separation within each sequence, and the absence 
of significant initial time-delay between the two sequences. 
Furthermore, we found that robust on-off and off-on switching 
of one of the propagating sequences can be achieved by an abrupt 
change in the ratio of cubic gain and quintic loss coefficients. 
Additionally, we demonstrated that the results can be extended 
to pulse sequences with periodically alternating phases. 
Our study significantly enhances the recently found relation 
between collision-induced dynamics of sequences of NLS solitons 
and evolution of population sizes in LV models. Moreover, it 
indicates that the relation might be further extended to sequences 
of solitary waves of the cubic-quintic GL equation, at least 
in some regions of parameter space. Finally, our results might 
find useful applications in nonlinear waveguides with saturable 
absorption for stable energy equalization and broadband switching.

\end{document}